# Hydrogen *s*-electrons as the origin of crystal magnetism beyond spin-orbit coupling

Baiqiang Liu[1], Zhen Gong[1], Rui Liu[1], Siyang Liu[1], Yue Feng[1], Zhigang Wang[1,2*]

[1] Key Laboratory of Material Simulation Methods & Software of Ministry of Education, College of Physics, Jilin University, Changchun 130012, China.
[2] Institute of Theoretical Chemistry, College of Chemistry, Jilin University, Changchun 130023, China.
[*]e-mail: wangzg@jlu.edu.cn (Z. W.)

**ABSTRACT**

Magnetism has long been attributed to localized *d*, *f*, and even *p* electrons with strong correlations, whereas *s* electrons exemplified by hydrogen are reactive and tend to have their spins quenched, making *s*-electron-derived magnetism and long-range ordered magnetic crystals seem unattainable. Here we report a low-Z ferromagnetic crystal $H_{13}@(BN)_{12}$ using first-principles calculations, where thirteen hydrogen atoms are encapsulated within a $(BN)_{12}$ cage and magnetism originates from the 1*s* electron of the central hydrogen atom. The crystal remains stable under ambient pressure owing to chemical precompression. Notably, the central hydrogen atom retains a magnetic moment of 1 μB, with long-range magnetic order established through multicenter bonding within the $H_{13}$ aggregate and the intercell B–B network, while the zero orbital angular momentum of *s* electrons renders spin-orbit coupling (SOC) negligible as expected. Electronic structure analyses reveal that the large cavity and central negative electrostatic potential of the $(BN)_{12}$ cage localize the hydrogen 1*s* electron, preventing spin quenching. Interestingly, under 16 GPa compression, the system transforms into a nonmagnetic metallic state driven by delocalized electrons of the central hydrogen atom. This study opens a pathway for constructing

*s*-electron-driven magnetic materials and lays the foundation for developing low-energy consumption magnetic devices without SOC.

## INTRODUCTION

Magnetic materials serve as cornerstones of cutting-edge technologies such as information storage, electronic systems, and quantum devices, with their development directly affecting various aspects of modern society [1-4]. Magnetism in conventional high-performance magnets mainly arises from *d/f* electrons [5-8], while *p* electrons can also contribute to magnetism in certain carbon-based systems [9-14]. The presence of orbital angular momentum in these electrons gives rise to spin-orbit coupling (SOC), which is directly associated with magnetic phenomena, including magnetic anisotropy, spin transport, and current-driven magnetization switching [15-21]. While SOC-mediated magnetization control enables diverse functionalities in current spintronic devices, it typically demands high current densities and suffers from substantial damping losses and Joule heating [22-26], thereby severely restricting its potential for low-power operation. The resulting energy bottleneck has become a key obstacle to the scalability and efficiency of next-generation magnetic materials. This situation raises a central question of whether robust magnetism can be realized in systems where SOC is intrinsically negligible.

Hydrogen *s* electrons possess zero orbital angular momentum, thereby offering a potential design pathway toward low-energy consumption magnetic materials. However, unlike *d/f* electrons, *s* electrons exhibit a strong tendency to form chemical bonds, making unpaired electrons difficult to stabilize and thereby hindering the formation of robust magnetic order [27,28]. As a result, no magnetic crystal driven by hydrogen *s* electrons has been realized to date, and the potential physical advantage arising from the absence of SOC remains unverified. Although some studies have attempted to induce *s*-electron magnetism in two-dimensional materials through defects or doping [29-32], the resulting moments are often unstable due to the nonuniform distribution of defects and the tendency of *s* electrons to pair with neighboring atoms, leading to the quenching of unpaired spins. Importantly, advances in understanding the behavior of hydrogen under confinement have opened up possibilities for localizing hydrogen *s* electrons and forming long-range ordered magnetic crystals [33,34]. Evidently, if robust magnetism could indeed arise from confined *s* electrons, although this remains unclear, it would enrich understanding of magnetism and enable a class of materials with a distinct electronic origin and low-energy consumption.

Against this background, we explore the feasibility of constructing magnetic crystals using hydrogen *s* electrons as the magnetic source. Through extensive structural modeling, we design a three-dimensional (3D) ferromagnetic (FM) crystal, $H_{13}@(BN)_{12}$, composed of low-Z elements and stabilized by chemical precompression under ambient pressure. Atomic-level analyses indicate that the 1*s* electron of the central hydrogen atom gives rise to magnetism, with magnetic moments exhibiting long-range ordering through multicenter bonding within the $H_{13}$ aggregate and the intercell B–B network, while SOC can indeed be ignored as predicted by fundamental physics. This behavior provides a pathway for low-energy consumption magnetic moment modulation. Notably, the large cavity of the $(BN)_{12}$ cage and its central negative electrostatic potential can localize the hydrogen 1*s* electron, thereby suppressing spin quenching. In addition, stress can modulate the system's magnetism, with tension concentrating the spin density on the central hydrogen atom and compression driving the system to a nonmagnetic (NM) metallic state induced by delocalized electrons of the central hydrogen atom. This work provides a design strategy for spintronic materials based on low-Z elements and has significant implications for both fundamental research and applications.

## RESULTS AND DISCUSSION

**Structural Features and Stability.** To explore intrinsic magnetism induced by hydrogen s electrons, we select the $(BN)_{12}$ cage as the host framework and construct a hydrogen-encapsulated system, inspired by hydrogen behavior under confinement [33,34]. Ultimately, we design an $H_{13}@(BN)_{12}$ system that remains stable under ambient pressure. As shown in Fig. 1a, nitrogen atoms form stable N–H bonds with hydrogen atoms, while the remaining one hydrogen atom resides at the center of the cage. The hydrogen atoms collectively form an approximately spherical structure with a radius of 1.42 Å. This feature arises from the large cavity of the $(BN)_{12}$ cage, which may promote the localization of the central hydrogen 1*s* electron, thereby providing a favorable electronic environment to prevent spin quenching and preserve the magnetic moment of the unpaired electron. For the $(BN)_{12}$ cage, N–H bond formation drives nitrogen atoms inward, while intercell bonding pushes the boron atoms outward, giving rise to the 3D crystal (Supplementary Fig. 1). The crystal belongs to the highly symmetric cubic space group *Pm-3* (No. 200), with an equilibrium lattice constant of 6.84 Å (detailed structural parameters are provided in Supplementary Table 1). Phonon dispersion calculations reveal no imaginary frequencies throughout the Brillouin zone (Fig. 1b), confirming the dynamical stability of the system under ambient pressure.

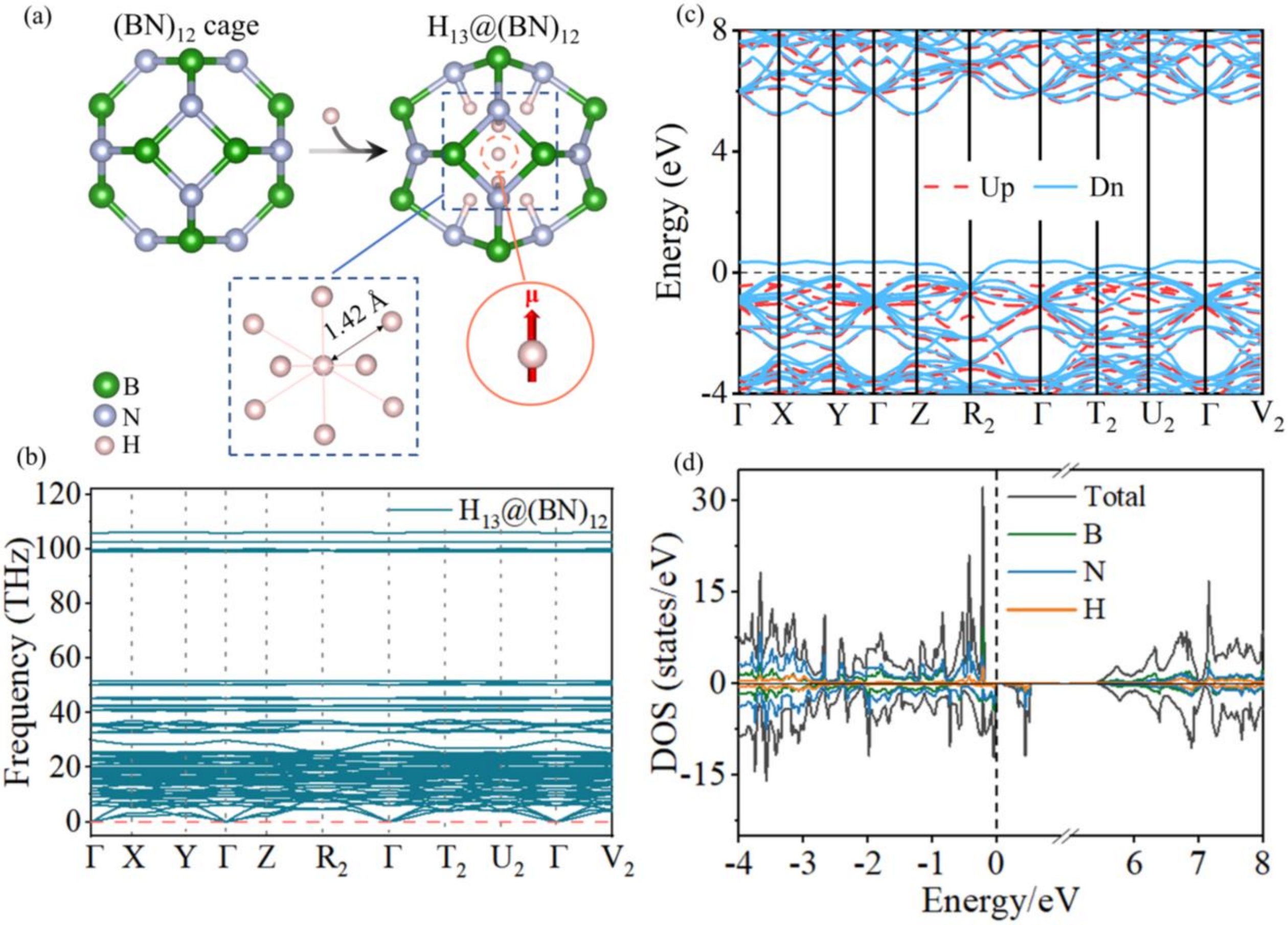

**Fig. 1 | Structural properties of the $H_{13}@(BN)_{12}$ crystal. a,** Schematic illustration of the primitive cell, showing hydrogen atoms encapsulated in the $(BN)_{12}$ cage. The blue rectangular region highlights the $H_{13}$ aggregate, which forms an overall nearly spherical structure with a radius of 1.42 Å. The red circled region indicates the magnetic moment associated with the central hydrogen 1*s* electron. **b,** Phonon dispersion curves. **c,d,** Band structure and DOS of the $H_{13}@(BN)_{12}$ crystal under ambient pressure. The high symmetry k points are Γ (0, 0, 0), X (0.5, 0, 0), Y (0, 0.5, 0), Z (0, 0, 0.5), $R_2$ (-0.5, -0.5, 0.5), $T_2$ (0, -0.5, 0.5), $U_2$ (-0.5, 0, 0.5), and $V_2$ (0.5, -0.5, 0). The Fermi level exhibits a distinct spin splitting.

Furthermore, we evaluate the mechanical stability of the $H_{13}@(BN)_{12}$ crystal by calculating elastic constants. The results are $C_{11}$ = 213.7 GPa, $C_{12}$ = 30.8 GPa, and $C_{44}$ = 24.4 GPa (Supplementary Table 2), all of which satisfy the Born-Huang criteria ($C_{11} > 0$, $C_{44} > 0$, and $C_{11} + 2C_{12} > 0$) [35], thereby confirming its mechanical robustness. Further first-principles molecular dynamics simulations indicate that the structure remains stable up to 40 K (Supplementary Fig. 2 and Fig. 3). Although this stability temperature is relatively low, it is still accessible under conventional cryogenic experimental conditions [36,37]. We also calculate the cohesive energy, which is 4.28 eV/atom (Supplementary Table 3), indicating strong bonding.

**Magnetic Properties and Stabilization Mechanism.** To elucidate the electronic structure of the $H_{13}@(BN)_{12}$ crystal, we perform band structure and density of states (DOS) calculations. As shown in Fig. 1c and 1d, the spin-resolved band structure and DOS calculated using the HSE06 hybrid functional reveal typical magnetic semimetallic properties [38,39], with clear spin splitting near the Fermi level. Electronic coupling occurs between the hydrogen atoms and the $(BN)_{12}$ cage, in which the spin-up channel displays a distinct band gap, while the spin-down channel crosses the Fermi level and exhibits metallic behavior. Furthermore, the PBE functional results exhibit the same behavior, reinforcing the reliability of the magnetic properties (Supplementary Fig. 4 and Fig. 5). Notably, this magnetism differs from that induced by defects or doping, as it originates from the intrinsic electronic structure of the system, thereby avoiding issues such as inhomogeneous magnetic-moment distribution and poor stability. These results demonstrate that magnetism can be realized in 3D low-Z crystals without relying on traditional magnetic elements.

In order to uncover the origin of this magnetic structure, it is necessary to investigate atomic-level features including bonding patterns, interaction mechanisms, and spin-density distributions. The electron localization function (ELF) is employed to examine the bonding properties [40]. As shown in Fig. 2a, electron localization is observed at both N–H and B–B bonds, indicating the formation of covalent bonds. Notably, the 1*s* electron of the central hydrogen atom becomes highly localized, contrary to the conventional expectation that *s* electrons are highly reactive and tend to undergo spin pairing. To elucidate this behavior, we perform an electrostatic potential analysis. As shown in Fig. 2b, removal of the central hydrogen atom yields the $H_{12}@(BN)_{12}$ system, where the strong polarity of the $(BN)_{12}$ cage generates a pronounced negative electrostatic potential at the cage center. Meanwhile, the large cavity of the $(BN)_{12}$ cage keeps the central hydrogen isolated from surrounding hydrogen atoms. Consequently, under the combined effect of the negative electrostatic potential well and the large cavity, the 1*s* electron of the central hydrogen atom contracts toward the cage center and forms a stable localized state.

To elucidate the interaction between the hydrogen aggregate and the $(BN)_{12}$ cage, as well as the resulting crystal properties, we perform a crystal orbital Hamilton population (COHP) analysis [41].

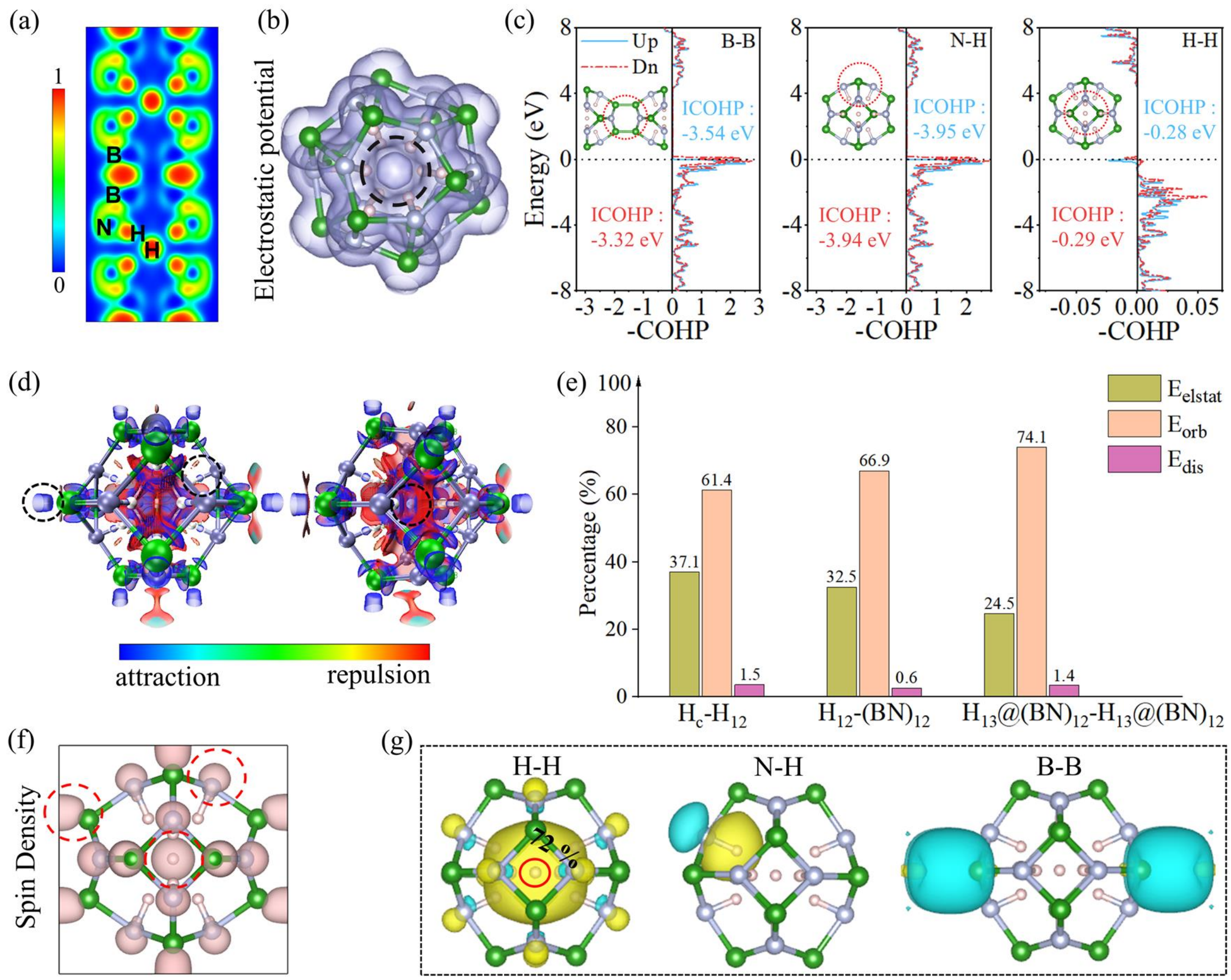

**Fig. 2 | Interaction analyses and magnetic properties of $H_{13}@(BN)_{12}$. a,** ELF map. ELF values close to 0 (blue) denote regions of delocalized electrons, whereas values close to 1 (red) indicate high electron localization. **b,** Electrostatic potential analysis. The bluish-purple region represents negative electrostatic potential. **c,** COHP and ICOHP analyses. **d,** IRI analysis. The blue, green, and red isosurfaces correspond to attractive, van der Waals, and repulsive interactions, respectively. **e,** EDA. Here, $H_c$ represents the center hydrogen atom, and $H_{12}$ represents the sphere surface hydrogen atoms. $E_{elstat}$ represents electrostatic interaction, $E_{orb}$ represents orbital interaction, and $E_{disp}$ represents dispersion interaction. **f,** Visualization of the spatial distribution of magnetic moments. The isosurfaces are set to 0.002 e/Å³. **g,** LMO analysis.

As shown in Fig. 2c, the B–B bonds exhibit dominant bonding contributions below the Fermi level, with spin asymmetry indicating spin polarization between primitive cells. In contrast, the N–H and H–H interactions show only weak spin splitting near the Fermi level, indicating a limited role in spin polarization. Notably, the N–H bonds exhibit strong bonding properties. Furthermore, one hydrogen atom is stably located at the center of the cage. Although its distance from the surrounding hydrogen atoms exceeds typical H–H bond lengths, a weak attractive interaction still exists. These results are also supported by the interaction region indicator (IRI) analysis [42]. As shown in Fig. 2d, the hydrogen atoms on the spherical surface exert attractive interactions on the central hydrogen atom, while the $(BN)_{12}$ cage imposes spatial steric constraints. This synergistic effect stabilizes the central hydrogen atom. Moreover, attractive interactions are also observed in both N–H and B–B bonds. These results are consistent with the chemical precompression effect [43], indicating that the stability of the $H_{13}$ aggregate originates from its chemical bonding interactions with the $(BN)_{12}$ cage.

To gain insight into the interaction mechanism from the perspective of energy contributions, we conduct energy decomposition analysis (EDA) [44]. As shown in Fig. 2e, the attraction between the central hydrogen atom and the surrounding hydrogen atoms is primarily governed by orbital interactions (61.4%). Attractive interactions between the hydrogen atoms on the spherical surface and the $(BN)_{12}$ cage, as well as between adjacent primitive cells, are also dominated by orbital interactions. Taken together with the ELF results, the COHP analysis confirms that $H_{13}@(BN)_{12}$ is a covalent crystal.

After elucidating the system's stability mechanism, we compare the relative energies of NM, FM, and antiferromagnetic (AFM) configurations to confirm the intrinsic magnetism of the $H_{13}@(BN)_{12}$ crystal. The results show that the FM state is the ground state, with energies lower than the NM and AFM states by 0.03 eV and 0.023 eV, respectively (Supplementary Table 4). Meanwhile, we examine stripe- and Néel-type AFM configurations in two dimensions, as well as A-, C-, and G-type AFM configurations in three dimensions, along with their corresponding FM states (Supplementary Table 5 and Table 6). These results consistently show that the FM configuration has the lowest energy, thereby confirming that $H_{13}@(BN)_{12}$ is a ferromagnetic crystal.

We also perform a spin-density analysis to investigate the microscopic origin of magnetism. As shown in Fig. 2f, the spin density is distributed over the boron, nitrogen, and central hydrogen atoms. The magnetic moment is thus extended across the entire system. Quantitative analysis of the magnetic moment corroborates this result, showing a total moment of 1 μB/unit arising from the central hydrogen atom and the $(BN)_{12}$ cage (Supplementary Table 7). Moreover, we examine the potential impact of SOC. The calculations show that the orbital magnetic moment is nearly quenched and the total magnetic moment remains unchanged upon inclusion of SOC (Supplementary Table 8 and Table 9), confirming negligible SOC in this system. Meanwhile, because the central hydrogen atom is located at a high-symmetry site at the center of the unit cell, all spatial directions are structurally equivalent. Combined with the analysis of the magnetic anisotropy energy, showing negligible energy differences among different magnetization directions (Supplementary Table 10), these results demonstrate that the system exhibits isotropic magnetic behavior. Notably, the spatially extended distribution of the magnetic moment differs from conventional system in which the magnetic moment is localized solely on the hydrogen atom [31,32]. This phenomenon highlights the necessity for further investigation.

To elucidate the formation mechanism of the magnetic distribution, localized molecular orbital (LMO) analysis is conducted. As shown in Fig. 2g, the $H_{13}@(BN)_{12}$ crystal features N–H and B–B bonds. More importantly, the $H_{13}$ aggregate shows multicenter bonding, providing channels for spin polarization. As a result, the magnetic moment is not confined to the central hydrogen atom but couples through multicenter bonding with the $(BN)_{12}$ cage, thereby extending the spin density to boron and nitrogen atoms. This result is corroborated by the above DOS analysis, which shows coupling between hydrogen and the $(BN)_{12}$ cage near the Fermi level, with the spin splitting contributed by both. Consequently, the magnetic moment exhibits a spatially extended distribution rather than following the conventional localized-moment model. More importantly, the B–B bonds that connect adjacent primitive cells provide a direct pathway for enabling the long-range distribution of spin polarization. The above COHP analysis also confirms this behavior, showing that the pronounced spin asymmetry of the B–B bonds not only contributes to the transmission of spin polarization but also plays an essential role in the magnetic coupling between primitive cells. Meanwhile, the spin-density distribution on the boron atoms further demonstrates that the B–B network effectively accommodates and transmit the spin polarization induced by the central hydrogen atom. Consequently, the magnetic moment originating from the central hydrogen is first coupled to the $(BN)_{12}$ cage through the multicenter bonding within the $H_{13}$ aggregate, and subsequently extends and couples between neighboring primitive cells via the spin-polarized B–B bond network, thereby establishing a robust long-range ferromagnetic order throughout the crystal.

**Comparative Structural Analyses and Nonmagnetic Behavior**. To clarify the role of the central hydrogen atom in magnetism, we investigate the $H_{12}@(BN)_{12}$ system, in which the central hydrogen is removed (detailed structural parameters are provided in Supplementary Table 11). Phonon dispersion calculations reveal no imaginary frequencies in the first Brillouin zone (Supplementary Fig. 6), confirming its dynamical stability. Both the band structure and DOS analyses demonstrate that this system is NM (Supplementary Fig. 7 and Fig. 8). Meanwhile, we also examine the structurally similar $H_{13}@C_{24}$ crystal (detailed structural parameters are provided in Supplementary Table 12), which likewise exhibits an NM ground state (Supplementary Fig. 9 and Fig. 10). These results collectively demonstrate the critical role of the $(BN)_{12}$ cage in localizing the 1*s* electron of the central hydrogen atom.

Unlike conventional magnetic materials that rely on *d/f* electrons, our study demonstrates that magnetic moments can originate solely from hydrogen *s* electrons and exhibit long-range spatial ordering. More importantly, in electrically controlled magnetization switching, conventional magnetic materials based on heavy elements typically require large currents for spin reversal, inevitably causing Joule heating [45,46]. In contrast, the magnetism in our system originates from hydrogen *s* electrons, with negligible SOC, potentially enabling more energy-efficient magnetization switching and low-energy consumption magnetic manipulation. This highlights the potential advantages of *s*-electron-dominated magnetic systems for future spintronic applications.

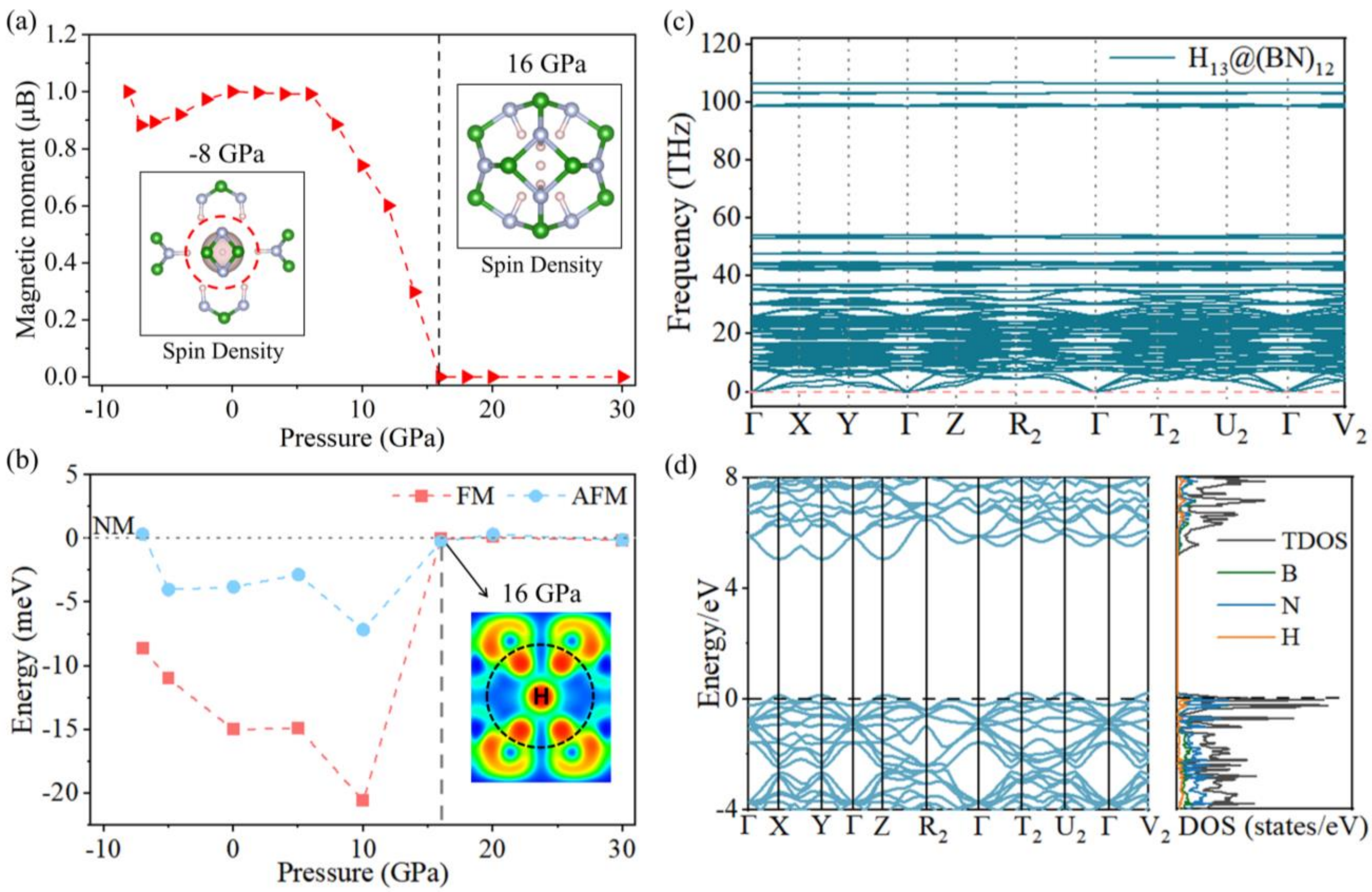

**Fig. 3 | Magnetic properties of the $H_{13}$@$(BN)_{12}$ crystal and their modulation under applied stress. a,** Evolution of the total magnetic moment. Under 8 GPa tensile stress, the $(BN)_{12}$ cage collapses, the spin density becomes localized on the central hydrogen atom, while under 16 GPa compressive stress, the system transitions from the FM state to the NM state. **b,** Energy comparison of different magnetic configurations. The energetic advantage of the FM state over the NM state vanishes under 16 GPa compressive stress. Here, the energy of the NM state is set as the reference (zero), and the energies of the FM and AFM states are shown as relative values. **c,** Phonon dispersion curves. **d,** Band structure and DOS of the $H_{13}$@$(BN)_{12}$ crystal under 16 GPa compressive stress.

Although *s* electrons are highly reactive and tend to overlap in solids, the $H_{13}$@$(BN)_{12}$ system benefits from the geometry of the $(BN)_{12}$ cage and the negative electrostatic potential at the cage center, which constraints the 1*s* electron of the central hydrogen atom mainly to localized states. Moreover, the multicenter bonding analysis shows that the contribution of the central hydrogen atom is approximately 72%, indicating that its electron remains largely localized around it. This localized environment ensures weak overlap of *s* electrons between adjacent primitive cells, preventing magnetic moment quenching and allowing Pauli exchange to dominate [47,48], thereby stabilizing the FM ground state.

**Stress-Induced Modulation of Magnetic Behavior.** Given the critical role of stress in modulating material properties, we investigate the effect of tensile and compressive stress on lattice stability and magnetism. As shown in Fig. 3a, tensile stress slightly reduces the total magnetic moment due to weakened interaction between the multicenter bonding within the $H_{13}$ aggregate and the $(BN)_{12}$ cage. At 8 GPa, the $(BN)_{12}$ cage fractures, causing the spin density that is distributed throughout the system under ambient pressure to become fully localized on the central hydrogen atom. In contrast, compressive stress decreases the magnetic moment linearly and quenches it at 16 GPa, indicating a transition from the FM to the NM states. Despite lattice contraction, the structure retains the *Pm-3* symmetry, confirming no structural phase transition (Supplementary Fig. 11). The energy comparison in Fig. 3b further supports the influence of stress on the magnetic configurations. Under tension, the FM state remains energetically more favorable than the NM state before the structural collapse occurs, whereas under compression, this advantage disappears at 16 GPa, corresponding to the quenching of magnetism.

To elucidate the origin of magnetic quenching, we perform an ELF analysis. The results reveal that the 1*s* electron localized on the central hydrogen atom under ambient pressure become delocalized under 16 GPa compressive stress (Fig. 3b). Instead, continuous yellow-green regions (ELF ≈ 0.5) emerge around the central hydrogen atom, while the typical deep-blue isolated regions between atoms vanish. This result indicates that magnetic quenching originates from compressive-stress-induced lattice contraction, which shortens H–H distances, enhances orbital overlap between central and surface hydrogen atoms, ultimately leading to the vanishing of net spin polarization and suppression of magnetism.

We conduct phonon dispersion calculations to verify the structural stability after magnetism quenching. As shown in Fig. 3c, no imaginary frequencies appear throughout the Brillouin zone, indicating dynamically stable. Further band structure and DOS calculations based on the HSE06 hybrid functional reveal that the compressive-stress-induced electronic reconstruction drives the $H_{13}$@$(BN)_{12}$ crystal to transform from a FM semimetallic state under ambient pressure into a NM metallic state (Fig. 3d), which originates from the delocalization of the central hydrogen electron. Overall, compared with conventional magnetic materials based on transition metals, the magnetism of the $H_{13}$@$(BN)_{12}$ crystal exhibits strong sensitivity to stress and can be completely suppressed under moderate compression, a property that may be experimentally detected and even modulated in the future.

To provide distinct spectroscopic features for potential experimental synthesis, we perform $^{1}$H nuclear magnetic resonance ($^{1}$H-NMR) calculations [49]. The results show that two distinct types of hydrogen atoms are present in $H_{13}$@$(BN)_{12}$, with an intensity ratio of approximately 12:1 (Supplementary Fig. 12), corresponding to the twelve equivalent hydrogen atoms bonded to nitrogen and one central hydrogen atom. Meanwhile, we perform vibrational analysis. The vibrational characteristics can be categorized into three regions, where the low-frequency region corresponds to the collective vibrations of the $H_{13}@(BN)_{12}$ system, the mid-frequency region arises mainly from the $H_{13}$ aggregate, and the high-frequency modes originate from hydrogen atoms bonded to nitrogen (Supplementary Fig. 13). These results provide clear spectroscopic fingerprints for future experimental identification.

**CONCLUSION**

In conclusion, we propose a strategy of encapsulating hydrogen atoms into $(BN)_{12}$ cages and utilizing the chemical precompression to construct a FM 3D crystal under ambient pressure, exhibiting magnetism originating from hydrogen *s* electrons. This approach offers four advantages: low-Z elements, magnetism induced by the hydrogen *s* electron with negligible SOC, long-range ordering of the magnetic moment, and stress-tunable magnetism. Specifically, hydrogen atoms bond to nitrogen atoms, inducing cage distortions that facilitate crystal formation. More importantly, the single hydrogen atom stabilized at the cage center serves as the origin of magnetism, and its magnetic moment is not localized on this hydrogen atom itself but instead exhibits a long-range ordered spatial distribution, with negligible influence from SOC.

Electronic structure analyses reveal that the large cavity of the $(BN)_{12}$ cage and its negative electrostatic potential localize the central hydrogen 1*s* electron, thereby suppressing *s*-electron overlap between adjacent primitive cells and stabilizing the FM ground state. We have also analyzed the electronic properties of the structurally similar $H_{12}@(BN)_{12}$ and $H_{13}@C_{24}$ crystals, and the results show that both systems exhibit NM behavior. This comparison further highlights the critical role of the $(BN)_{12}$ cage in enabling hydrogen *s*-electron magnetism. Moreover, the magnetism of the $H_{13}@(BN)_{12}$ crystal is highly sensitive to applied stress. Tensile stress causes the originally extended spin density to contract and localize on the central hydrogen atom, whereas compressive stress induces delocalization of the central hydrogen electron, thereby driving the system to transition from a FM semimetallic state under ambient pressure to a NM metallic state. Spectroscopic calculations further provide guidance for experimental identification. This work breaks through the conventional paradigm in which magnetism originates from *d*/*f* electrons, and establishes a platform for low-Z spintronic materials.

## METHODS

The calculations and simulations were performed using the Vienna Ab initio Simulation Package (VASP) based on density functional theory (DFT) [50]. The Perdew-Burke-Ernzerhof (PBE) functional within the generalized gradient approximation (GGA) was used to describe the exchange-correlation interactions [51]. During structural optimization, the lattice parameters and atomic positions were fully relaxed until the force on each atom was less than $10^{-4}$ eV/Å, with the energy convergence criterion set at $10^{-8}$ eV. Electronic structure calculations were performed using a plane-wave cutoff energy of 800 eV, together with a 5 × 5 × 5 Monkhorst-Pack k-point mesh for Brillouin-zone sampling. First-principles molecular dynamics simulations were performed using the CP2K software package [52], employing a 2×2×2 supercell with a time step of 0.5 fs. The system reached equilibrium after 5 ps, followed by subsequent 20 ps simulations. To verify the stability of the hydrogen-rich compound crystals, the PHONOPY program was used to calculate the phonon dispersion of the crystals [53]. Crystal orbital Hamilton population (COHP) and integrated COHP (ICOHP) calculations were performed using the LOBSTER package [41,54]. Interaction region indicator (IRI) [42], energy decomposition analysis (EDA) [44], and local molecular orbital (LMO) analyses were conducted using the Multiwfn program [55]. The DFT-D3 method was included in the IRI and EDA analyses [56]. In the EDA analysis, we mainly focused on the attractive interactions.

Therefore, only the electrostatic, orbital, and dispersion contributions were considered. The nuclear magnetic resonance (NMR) analysis was carried out using the Castep software [49,57]. We also calculated the cohesive energy to assess the stability of the crystal from an energetic perspective, and its expression is given as follows:

$$E_{\mathrm{coh}} = \frac{\sum E_{\mathrm{atom}} - E_{\mathrm{system}}}{N}$$

Here, $E_{\mathrm{atom}}$ represents the energies of the N individual atoms, and $E_{\mathrm{system}}$ denotes the total energy of the system.

## ACKNOWLEDGMENT

This work was supported by the National Key Research and Development Program of China (No. 2024YFA1409900), the Science and Technology Development Program of Jilin Province of China (20250102014JC) and the National Natural Science Foundation of China (grant number 11974136).

**AUTHOR CONTRIBUTIONS**

Z. W. initiated and supervised the work. B. L. performed simulations and calculations. B. L, Z. G, R. L, S. L., Y. F. and Z. W. discussed the results. B. L. and Z. W. prepared the manuscript.